\documentclass[conference]{IEEEtran}
\IEEEoverridecommandlockouts
\usepackage{cite}
\usepackage{amsmath,amssymb,amsfonts}
\usepackage{algorithmic}
\usepackage{graphicx}
\usepackage{textcomp}
\usepackage{xcolor}

\usepackage{amsthm}
\usepackage{bbm}
\usepackage{stmaryrd}
\usepackage{epic,eepic}

\usepackage{epsf}           
\usepackage{graphicx}       
\usepackage{epsfig}         
\usepackage{pstricks}
\usepackage{pstricks-add}
\usepackage{pst-plot}
\usepackage{dsfont}
\usepackage{hypernat}
\usepackage{hyperref}

\newtheorem{theorem}{Theorem}
\newtheorem{remark}{Remark}
\newtheorem{definition}{Definition}

\newtheorem{lemma}{Lemma}

\newcommand{\mc}{\mathcal}

\newcommand{\PP}{\mathcal{P}}

\newcommand{\QQ}{\mathcal{Q}}
\newcommand{\R}{\mathbb{R}}

\newcommand{\N}{\mathbb{N}}
\newcommand{\E}{\mathbb{E}}
\newcommand{\Q}{\mathbb{Q}}

\definecolor{darkgreen}{rgb}{0.0,0.6,0}

\def\BibTeX{{\rm B\kern-.05em{\sc i\kern-.025em b}\kern-.08em
    T\kern-.1667em\lower.7ex\hbox{E}\kern-.125emX}}

\usepackage[nomessages]{fp}
\newcommand\const[3][6]{%
    \edef\temporary{round(#3}%
    \expandafter\FPeval\csname#2\expandafter\endcsname
        \expandafter{\temporary:#1)}%
        \pstVerb{/#2 \csname#2\endcsname\space def}%
}

\newcommand{\FigMMSEbis}[2]{

\const{Pa}{0.5 * (#1 - 2*#2 - (#1*(#1- 4*#2))^0.5)}
\const{Pb}{0.5 * (#1 - 2*#2 + (#1*(#1- 4*#2))^0.5)}

\const{MMSEzero}{#1 * #2 / (#1 + #2)}

\const{MaxOrd}{\MMSEzero * 1.1}
\const{MaxAbs}{#1 * 1.1}

\const{MinOrd}{-0.1 * \MaxOrd}
\const{MinAbs}{-0.1 * \MaxAbs}

\const{XUNIT}{1/\MaxAbs * 7}
\const{YUNIT}{1/\MaxOrd * 5}

\const{AbsLin}{#1- #2}
\const{OrdLin}{#2/#1 * (#1 - #2)}

\def\MMSEsa{ #1 sqrt x sqrt neg add dup mul #2 mul #1 sqrt x sqrt neg add dup mul #2 add div}
\def\MMSElin{#1 x neg add  #2 neg add #2 mul #1 div}

\const{MMSEPa}{#2 /(2*#1) * (#1 + (#1*(#1-4*#2))^0.5)}
\const{MMSEPb}{#2 /(2*#1) * (#1- (#1*(#1-4*#2))^0.5)}

\const{MinOrd}{-0.02}

\begin{figure}[!h]
\begin{center}
\psset{xunit= \XUNIT cm,yunit= \YUNIT  cm}
\begin{pspicture}(-0.1,\MinOrd)(\MaxAbs,\MaxOrd)
\psline{->}(0,\MinOrd)(0,\MaxOrd)
\psline{->}(\MinAbs,0)(\MaxAbs,0)
\psplot[plotpoints=200,linecolor=blue]{0}{#1}{ \MMSEsa}
\psplot[plotpoints=200,linecolor=brown]{0}{\AbsLin}{\MMSElin}

\rput[r](-0.03,\MinOrd){$0$}
\rput[u](\MaxAbs,\MinOrd){$P$}
\rput[r](-0.03,\MaxOrd ){$S$}
\rput[u](#1,\MinOrd){$Q$}

\rput[u](#1,\MinOrd){$Q$}

\rput[u](0.1,\MinOrd){$P_1$}
\rput[u](\Pb,\MinOrd){$P_2$}

\rput[u](0.93,0.042){$Q=#1$, $N=#2$}

\psline[linestyle=dotted](\Pa,0)(\Pa,\OrdLin)
\psline[linestyle=dotted](\Pb,0)(\Pb,\OrdLin)
\psline[linestyle=dotted](0,\MMSEPa)(\AbsLin,\MMSEPa)
\psline[linestyle=dotted](0,\MMSEPb)(\AbsLin,\MMSEPb)

\psdots(\AbsLin,0)(#1,0)(\Pa,0)(\Pb,0)
\psdots(0,\OrdLin)(0,\MMSEzero)(0,\MMSEPa)(0,\MMSEPb)

\end{pspicture}
\caption{The curve $S_{\ell}(P)$ in \eqref{eq:MMSElinear} and the straight line $ \frac{N\cdot(Q-N-P)}{Q} $.}\label{fig:OptimalMMSE_#1_#2}
\end{center}
\end{figure}}

\newcommand{\FigMMSEThird}[2]{

\const{Pa}{0.5 * (#1 - 2*#2 - (#1*(#1- 4*#2))^0.5)}
\const{Pb}{0.5 * (#1 - 2*#2 + (#1*(#1- 4*#2))^0.5)}

\const{MMSEzero}{#1 * #2 / (#1 + #2)}

\const{MaxOrd}{\MMSEzero * 1.1}
\const{MaxAbs}{#1 * 1.1}

\const{MinOrd}{-0.1 * \MaxOrd}
\const{MinAbs}{-0.1 * \MaxAbs}

\const{XUNIT}{1/\MaxAbs * 7}
\const{YUNIT}{1/\MaxOrd * 5}

\const{AbsLin}{#1- #2}
\const{OrdLin}{#2/#1 * (#1 - #2)}

\def\MMSEsa{ #1 sqrt x sqrt neg add dup mul #2 mul #1 sqrt x sqrt neg add dup mul #2 add div}
\def\MMSElin{#1 x neg add  #2 neg add #2 mul #1 div}

\const{MMSEPa}{#2 /(2*#1) * (#1 + (#1*(#1-4*#2))^0.5)}
\const{MMSEPb}{#2 /(2*#1) * (#1- (#1*(#1-4*#2))^0.5)}

\const{MinOrd}{-0.07}
\const{OrdConstant}{0.75}

\begin{figure}[!h]
\begin{center}
\psset{xunit= \XUNIT cm,yunit= \YUNIT  cm}
\begin{pspicture}(-0.5,-0.1)(\MaxAbs,\MaxOrd)
\psline{->}(0,\MinOrd)(0,\MaxOrd)
\psline{->}(\MinAbs,0)(\MaxAbs,0)
\psplot[plotpoints=200,linecolor=blue]{0}{#1}{ \MMSEsa}
\psplot[plotpoints=200,linecolor=brown]{0}{\AbsLin}{\MMSElin}

\rput[r](-0.2,\MinOrd){$0$}
\rput[u](\MaxAbs,\MinOrd){$P$}
\rput[r](-0.1,\MaxOrd ){$S$}
\rput[u](#1,\MinOrd){$Q$}

\rput[u](#1,\MinOrd){$Q$}

\rput[l](\Pa,\MinOrd){$P_1$}
\rput[u](\Pb,\MinOrd){$P_2$}

\rput[u](\AbsLin,\OrdConstant){$Q=#1$, $N=#2$}

\psline[linestyle=dotted](\Pa,0)(\Pa,\OrdLin)
\psline[linestyle=dotted](\Pb,0)(\Pb,\OrdLin)
\psline[linestyle=dotted](0,\MMSEPa)(\AbsLin,\MMSEPa)
\psline[linestyle=dotted](0,\MMSEPb)(\AbsLin,\MMSEPb)

\psdots(\AbsLin,0)(#1,0)(\Pa,0)(\Pb,0)
\psdots(0,\OrdLin)(0,\MMSEzero)(0,\MMSEPa)(0,\MMSEPb)

\fileplot[linecolor=darkgreen]{dataTwoPoint/DataTwoPoint_10_1.dat}

\end{pspicture}
\caption{The lasso-shaped curve corresponds to the Witsenhausen's two-point strategy with parametric equations \eqref{eq:TwoPointP} and \eqref{eq:TwoPointS} for  $a\in[0.05,5]$. }\label{fig:OptimalMMSE_#1_#2_bis}
\end{center}
\end{figure}}

\begin{document}

\title{
Continuous Random Variable Estimation is not Optimal for the Witsenhausen Counterexample
\thanks{
The authors gratefully acknowledge the financial support of SRV ENSEA for visits at KTH in Stockholm in 2017 and 2019, and at ETIS in Cergy in 2018. This research has been conducted as part of the Labex MME-DII (ANR11-LBX-0023-01). Part of the research has been supported by Swedish Research Council (VR) under grant 2020-03884.
}
}

\author{\IEEEauthorblockN{Ma\"{e}l Le~Treust}
\IEEEauthorblockA{\textit{ETIS UMR 8051,} \\
\textit{CY Cergy Paris Universit\'{e}, ENSEA, CNRS}\\
95014 Cergy-Pontoise CEDEX, France 
}
\and
\IEEEauthorblockN{Tobias J. Oechtering}
\IEEEauthorblockA{\textit{KTH Royal Institute of Technology EECS} \\
\textit{Div. Inform. Science and Engineering}\\
10044 Stockholm, Sweden 
}
}

\maketitle

\begin{abstract}
Optimal design of distributed decision policies can be a difficult task, illustrated by the famous Witsenhausen counterexample. In this paper we characterize the optimal control designs for the vector-valued setting assuming that it results in an internal state that can be described by a continuous random variable which has a  probability density function. More specifically, we provide a genie-aided outer bound that relies on our previous results for empirical coordination problems. This solution turns out to be not optimal in general, since it consists of a time-sharing strategy between two linear schemes of specific power. It follows that the optimal decision strategy for the original scalar Witsenhausen problem must lead to an internal state that cannot be described by a continuous random variable which has a probability density function.

\end{abstract}



\section{Introduction}
Distributed decision-making systems arise in many engineering problems where decentralized agents choose actions based on locally available information as to minimize a common cost function. The information at each agent is either locally observed or received from other agents. Since the process of sharing information comes with a cost, agents usually do not have access to the whole information 
available at all agents. The design of optimal decision strategies for such problems 
is considered to be notoriously difficult. The Witsenhausen counterexample \cite{W68acis} from 1968  
is an outstanding toy example that has significantly helped to understand the fundamental 
difficulty that actions serve two purposes, a control purpose affecting the system 
state and a communication purpose providing information to other agents \cite{YB13sncs}.

Although Witsenhausen refuted with his simple two-point counterexample the assertion that a linear policy would be also optimal in such a Gaussian setting, the optimal non-linear policy remains unknown. Many researcher have approached the optimization problem with various methods. In the last decade for instance it has been approached with numerical optimization methods \cite{TT17alsa},\cite{KGOS11iscc}, where the latter is based on an iterative source-channel coding approach. Analytically,
using results from optimal transport theory, it has been shown in \cite{WV11wcav} 
that the optimal decision strategy is a strictly increasing unbounded piece-wise real analytic function with a real analytic left inverse. 
More necessary conditions have been derived in \cite{MH15ofam} by analyzing 
an equivalent optimization problem on the space of square-integrable quantile functions. 
However, it is unclear if the optimal decision policy of the first agent results in an internal state that can be described by a continuous random variable.

In this work, we show that the optimal decision strategy will not lead to an internal state that can be described by a continuous random variable that has a probability density function.
The observation points on a subtle point in an outer bound argument, which might be easily overseen. We will further discuss that this observation, and in essence also the Witsenhausen counterexample, can be easily explained by the relation between the MMSE and mutual information considering Gaussian or binary distributed input \cite{GSV05miam}. 

\begin{figure}[!ht]
\begin{center}
\psset{xunit=1cm,yunit=1cm}
\begin{pspicture}(0.5,-0.4)(12,1.8)
\psframe(2,0)(3,1)

\pscircle(4.2,0.5){0.2}
\rput(4.2,0.5){$+$}
\pscircle(5.6,0.5){0.2}
\rput(5.6,0.5){$+$}
\psframe(6.8,0)(7.8,1)

\psdots[linewidth=2pt](1,1.5)(5.6,1.5)
\rput[u](2.4,1.8){$X_0^{n}\sim \mathcal{N}(0,Q\mathbb{I})$}
\psline[linewidth=1pt]{->}(1,1.5)(1,0.5)(2,0.5)
\psline[linewidth=1pt]{->}(1,1.5)(4.2,1.5)(4.2,0.7)
\psline[linewidth=1pt]{->}(3,0.5)(4,0.5)
\psline[linewidth=1pt]{->}(4.4,0.5)(5.4,0.5)
\psline[linewidth=1pt]{->}(4.9,0.5)(4.9,-0.5)(8.8,-0.5)
\psline[linewidth=1pt]{->}(5.6,1.5)(5.6,0.7)
\psline[linewidth=1pt]{->}(5.8,0.5)(6.8,0.5)
\psline[linewidth=1pt]{->}(7.8,0.5)(8.8,0.5)

\rput[u](1.5,0.8){$X_0^n$}
\rput[u](3.5,0.8){$U_1^n$}
\rput[u](4.9,0.8){$X_1^{n}$}
\rput[u](6.3,0.8){$Y_1^{t}$}
\rput[u](6,1.8){$Z_1^{n}\sim \mathcal{N}(0,N\mathbb{I})$}
\rput[u](8.3,0.8){$U_{2,t}$}
\rput[u](8.3,-0.2){$X_{1,t}$}
\rput(2.5,0.5){$C_1$}
\rput(7.3,0.5){$C_2$}

\end{pspicture}
\caption{The state and the channel noise are drawn according to the i.i.d. Gaussian distributions $X_0^{n}\sim \mathcal{N}(0,Q\mathbb{I})$ and $Z_1^{n}\sim \mathcal{N}(0,N\mathbb{I})$. The internal state sequence $X_1^n$ is causally estimated by decision maker $C_2$.}
\end{center}
\end{figure}
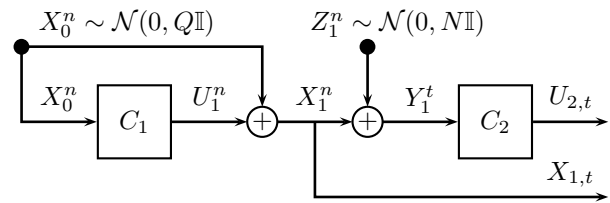

Our approach is based on a vector-valued extension of the Witsenhausen counterexample as proposed by Grover and Sahai in  \cite{GS10wcaa}. They study a non-causal encoding and decoding strategy that combines a coding scheme with side information and a linear scheme,
which has been shown to be optimal by Choudhuri and Mitra in \cite{CM12owct}. It has later been observed that such problems can be also approached as an empirical coordination coding problem. In  \cite{MO18ocdv}, we have provided an overview on the individual findings and completed the missing cases using coding results 
from \cite{MT17jeco}. In \cite{OL19cccd} we have derived an achievability result considering non-causal encoding and causal decoding using a continuous alphabet building on proof methods from \cite{VOS18hiwp}. In this work, we now derive a genie-aided outer bound for this case considering only decision strategies that result in continuous random variables which have a probability density functions.

\section{System Model}

In this work, we restrict our study to continuous random variables which have a probability density function (pdf), see \cite[Chap. 8]{CT06eoit}. For brevity we only refer to continuous random variables.

We consider the vector-valued Witsenhausen setup in which the sequences of states and channel noises are drawn independently according to the i.i.d. Gaussian distributions $X_0^{n}\sim \mathcal{N}(0,Q\mathbb{I})$ and $Z_1^{n}\sim \mathcal{N}(0,N\mathbb{I})$ with $\min(Q,N)>0$, where $\mathbb{I}$ denotes the identity matrix. We denote by $X_1$ the \emph{internal state} and $Y_1$ the output of the noisy channel.
\begin{align}
X_1 =& X_0 + U_1 \qquad\qquad\qquad\;\quad  \text{ with }X_0 \sim\;\; \mc{N}(0,Q),\label{eq:Gaussian1} \\
Y_1 =& X_1 + Z_1=X_0 + U_1+ Z_1  \text{ with }Z_1 \sim\;\; \mc{N}(0,N).\label{eq:Gaussian3}
\end{align}
We denote by $\PP_{X_0} = \mc{N}(0,Q)$ the Gaussian state distribution and by $\PP_{X_1Y_1|X_0U_1}$ the conditional probability distribution corresponding to equations \eqref{eq:Gaussian1} and \eqref{eq:Gaussian3}.

\begin{definition}\label{def:Code}
For $n\in\N^{\star} = \N \setminus\{0\}$, a ``control design'' with non-causal encoder and causal decoder is a tuple of stochastic functions $c=(f,\{g_t\}_{t\in\{1,\ldots,n\}})$ defined by
\begin{align}
f : \mc{X}_0^n  \longrightarrow   \mc{U}_1^n,\quad
g_t :  \mc{Y}_1^{t}  \longrightarrow \mc{U}_2 ,   \;\forall t \in \{1, \ldots,n\}, \label{eq:Code}
\end{align}
which induces a distribution over the sequences given by
\begin{align*}
&\bigg(\prod_{t=1}^n \PP_{X_{0,t}}\bigg)f_{U_1^n|X_0^n}\bigg(\prod_{i=t}^n \PP_{X_{1,t}Y_{1,t}|X_{0,t}U_{1,t}} \bigg)
\bigg( \prod_{t=1}^n g_{U_{2,t}|Y_1^{t}}\bigg).\label{eq:Distribution}
\end{align*}
We denote by $\mc{C}_{\textsf{d}}(n)$ the set of control designs with non-causal encoder and causal decoder $c=(f,\{g_t\}_{t\in\{1,\ldots,n\}})$ that induce sequences of continuous random variables.
\end{definition}

\begin{definition}
We define the $n$-stage costs associated with $c$ by
\begin{align}
\gamma^n_{\textsf{p}}(c) =&
\begin{cases}
 \E\Big[\frac{1}{n}\sum_{t=1}^n U_{1,t}^2\Big] &\text{if it exists,}\\
+\infty &\text{otherwise,}
\end{cases}\\
\gamma^n_{\textsf{s}}(c) =&\begin{cases}
 \E\Big[ \frac{1}{n}\sum_{t=1}^n (X_{1,t} - U_{2,t})^2\Big] &\text{if it exists,}\\
+\infty &\text{otherwise.}
\end{cases}
\end{align}
The pair of costs $(P,S)\in\R^2$ is achievable if for all $\varepsilon>0$, there exists $\bar{n}\in\N^{\star}$, for all $n\geq \bar{n}$, there exists a control design $c\in\mc{C}_{\textsf{d}}(n)$ such that
\begin{align}
\Big| P - \gamma^n_{\textsf{p}}(c) \Big| + 
\Big| S - \gamma^n_{\textsf{s}}(c) \Big| \leq \varepsilon.\label{eq:EqualConstraint}
\end{align}
\end{definition}

\begin{theorem}\label{theo:MainResult}
The pair of Witsenhausen costs $(P,S)$ is achievable if and only if  there exists continuous random variables with probability distribution that decomposes according to 
\begin{align}
&\PP_{X_0} \QQ_{U_1W_1W_2|X_0} \PP_{X_1Y_1|X_0U_1} \QQ_{U_2|W_2Y_1},\label{eq:TargetDistribution0}
\end{align}
where $(W_1,W_2)$ are auxiliary random variables such that $0\leq~I(W_1;Y_1|W_2) - I(W_1,W_2;X_0)$ and \begin{align}
P = \E_{\QQ}\big[U_{1}^2\big],\qquad
S = \E_{\QQ}\big[(X_{1} - U_{2})^2\big].
\end{align}
\end{theorem}
This result is stated in \cite[Theorem 1]{OL19cccd}.

\begin{remark}
The probability distribution in \eqref{eq:TargetDistribution0} satisfies
\begin{align}
\begin{cases}
(X_1,Y_1)  -\!\!\!\!\minuso\!\!\!\!- (X_0 ,U_1) -\!\!\!\!\minuso\!\!\!\!-  (W_1,W_2) ,\\ 
U_2 -\!\!\!\!\minuso\!\!\!\!- (Y_1 , W_2 ) -\!\!\!\!\minuso\!\!\!\!-  (X_0,X_1,U_1, W_1 ) , \\
X_2 -\!\!\!\!\minuso\!\!\!\!- (X_1,U_2 ) -\!\!\!\!\minuso\!\!\!\!-  (X_0,U_1,Y_1, W_1,W_2 ).
\end{cases}\label{eq:MarkovChains}
\end{align}
The causality condition prevents the controller $C_2$ to recover $W_1$ which induces the second Markov chain of \eqref{eq:MarkovChains}.
\end{remark}

\begin{definition}
The optimal cost considering continuous random variables is characterized by the optimization problem defined as follows
\begin{align}
&S_{\textsf{c}}(P) = \min_{\QQ \in \Q_{\textsf{c}}(P)}{\E_{\QQ}\big[(X_{1} - U_{2})^2\big]},\label{eq:MMSEContinuous} \\
&\Q_{\textsf{c}}(P) = \bigg\{ \big( \QQ_{U_1W_1W_2|X_0}, \QQ_{U_2|W_2Y_1}\big) \;\text{ s.t. }\; P = E_{\QQ}\big[U_{1}^2\big], \nonumber \\
&\;\; I(W_1;Y_1|W_2) - I(W_1,W_2;X_0)\geq0 \text{ and }\nonumber \\
&\;\; X_0,U_1,W_1,W_2,X_1,Y_1, U_2\;  \text{ are continuous }
  \bigg\}.\label{eq:InformationConstraintContinuous}
\end{align}
\end{definition}

\begin{lemma}[Lemma 11 in \cite{W68acis}]
The best linear scheme is $U_1 = -\sqrt{\frac{P}{Q}} X_0$ if $P\leq Q$, otherwise $U_1 = -X_0+\sqrt{P-Q}$.\footnote{If $P> Q$, the inernal state $X_1$ can be canceled and the offset $\sqrt{P-Q}$ 
is only included to meet the power constraint with equality as in \eqref{eq:EqualConstraint}.} This induces the estimation cost given by
\begin{align}
S_{\ell}(P)=& 
\begin{cases}
\frac{ \big(\sqrt{Q}-  \sqrt{P} \big)^2 \cdot N}{ \big(\sqrt{Q}-  \sqrt{P} \big)^2 + N}\quad &\text{ if } P\in[0,Q],\\
0 \qquad\qquad &\text{ otherwise.}
\end{cases}\label{eq:MMSElinear}
\end{align}
\end{lemma}

\FigMMSEbis{2}{0.2}

\begin{theorem}
\label{theo:CharacterizationContinuousRV}
The optimal cost with continuous random variables satisfies
\begin{align}
S_{\textsf{c}}(P) =&
\begin{cases}
\frac{N\cdot(Q-N-P)}{Q} & \text{if } Q> 4N \text{ and } P\in[P_1,P_2],\\
S_{\ell}(P) & \text{otherwise, }
\end{cases}
\label{eq:SolutionContinuous}\\
P_1 =& \frac{1}{2} \Big(Q - 2N - \sqrt{Q\cdot(Q-4N)}\Big),\\
P_2 =& \frac{1}{2} \Big(Q - 2N +\sqrt{Q\cdot(Q-4N)}\Big).
\end{align}
\end{theorem}

The proof of Theorem \ref{theo:CharacterizationContinuousRV} is stated in Sec. \ref{sec:ConverseProof} and \ref{sec:AchievabilityProof}. Figure \ref{fig:OptimalMMSE_2_0.2} represents the cost of the linear scheme $S_{\ell}(P)$ and the line of equation $P\mapsto  \frac{N\cdot(Q-N-P)}{Q} $. Note that the upper bound in \eqref{eq:SolutionContinuous} may be obtained by using either a time sharing strategy between the two linear schemes with parameters $P_1$ and $P_2$ when $Q> 4N$ and $P\in[P_1,P_2]$, or a linear scheme of power $P$. This result shows that memoryless policies are optimal so that these policies are also optimal for the original scalar Witsenhausen counterexample setup restricted to continuous random variables. However, as pointed out by Witsenhausen  in \cite{W68acis}, such a strategy in the original scalar model is generally not optimal!



\section{Witsenhausen's two-points strategy}\label{sec:WitsenhausenTwoPoints}

The two-point strategy described in \cite[pp. 141]{W68acis} outperforms the optimal cost with continuous random variables $S_{\textsf{c}}(P)$ of Theorem \ref{theo:CharacterizationContinuousRV}. We consider $Q=10$, $N=1$ and the sender's strategy with parameter $a\geq0$ given by
\begin{align}
U_1 =& a \cdot \textrm{sign}\big(X_0\big) - X_0,
\end{align}
which induces $X_1 = a \cdot \textrm{sign}\big(X_0\big)$. For all $a\geq0$, the pair of costs are given by
\begin{align}
P_{\,\textsf{t}}(a)=&Q + a\Big(a-2\sqrt{\frac{2Q}{\pi}}\Big),\label{eq:TwoPointP}\\
S_{\,\textsf{t}}(a) =&\frac{a^2}{\sqrt{2 \pi}}   e^{-\frac{a^2}{2}} \int \frac{ e^{-\frac{y^2}{2}}}{\cosh{a y}} dy.\label{eq:TwoPointS}
\end{align}
Fig. \ref{fig:OptimalMMSE_10_1_bis} shows that for some $a\in[0.05,5]$, the pair of costs $(P_{\,\textsf{t}}(a),S_{\,\textsf{t}}(a))$ Pareto-dominates $S_{\textsf{c}}(P)$ of Theorem \ref{theo:CharacterizationContinuousRV}, for some $P$.
\FigMMSEThird{10}{1}

From the previous, we have the following Theorem.
\begin{theorem}
There exists values for $Q$, $N$, and $a$ for which we have $S_{\,\textsf{t}}(a)\leq S_{\textsf{c}}(P_{\,\textsf{t}}(a))$.
\end{theorem}

\subsection*{Discussion}\label{subsec:Discussion}
In the following we also briefly want to explain the Witsenhausen counterexample result in terms of the I-MMSE formula by Guo, Shamai and Verdu \cite{GSV05miam}. The formula has been used to illustrate in \cite[Fig. 1]{GSV05miam} that binary inputs in additive Gaussian noise channels result in a lower MMSE than Gaussian distributed inputs with the same SNR. In other words, to achieve the same MMSE, binary distributed input requires less channel input power than Gaussian distributed input. 
Exactly this power gain has been exploited in the Witsenhausen counterexample scheme in which the internal state $X_1$ is binary so that the resulting MMSE outperforms the MMSE of the best linear scheme. Analytically, it is interesting to see that the MMSE formulas \eqref{eq:TwoPointP} and \eqref{eq:TwoPointS} directly relate to \cite[Equations (13) and (17)]{GSV05miam} 
with the corresponding signal powers and noise power.

\section{Conclusion}\label{sec:Conclusion}
Our results show that information theoretic methods, in particular coordination coding results, lead to new insights on the Witsenhausen counterexample. Vice versa, we believe that our observation makes the Witsenhausen counterexample also interesting for other source-channel coding problems. In more detail, we show that a convex combination of linear memoryless policies is optimal for the vector-valued Witsenhausen problem with causal decoder restricted to the space of continuous random variables. Since the policy is memoryless, it follows that the linear policy is also optimal for the original Witsenhausen problem restricted to the space of continuous random variables which have a pdf. However, Witsenhausen's two-points strategy outperforms the best linear strategy, which implies that the hypothesis of a continuous random variable which has a pdf is an active restriction for the  Witsenhausen counter-example setup. 
According to the Lebesgue's decomposition Theorem, every probability distribution on the real line is a mixture of discrete part, singular part, and an absolutely continuous part. Accordingly, we conclude that the optimal decision strategy for the unrestricted Witsenhausen's counter-example must lead to an internal state that cannot be described by a continuous random variable which has a pdf. In future works, we will consider policies that result in internal states described by more general probability distributions.

\section{Proofs}
\subsection{Lower bound for the Theorem \ref{theo:CharacterizationContinuousRV}}\label{sec:ConverseProof}

The Markov chain $Y_1 -\!\!\!\!\minuso\!\!\!\!- (X_0,U_1)  -\!\!\!\!\minuso\!\!\!\!- (W_1,W_2)$ implies
\begin{align}
&I(W_1;Y_1|W_2) - I(W_1,W_2;X_0) \nonumber\\
\leq& I(W_1;Y_1|W_2,X_0) - I(W_2;X_0)\\
\leq& I(U_1;Y_1|W_2,X_0) - I(W_2;X_0).
\end{align}
Therefore
\begin{align}
&S_{\textsf{c}}(P) 
\geq \min_{\QQ_{U_1W_2|X_0}  \in {\Q}_1(P),\atop \QQ_{U_2|W_2Y_1}}{\E_{\QQ}\big[(X_{1} - U_{2})^2\big]}\\
\geq& \min_{\QQ_{U_1W_2|X_0}  \in \Q_1(P)} \E_{\QQ}\Big[\Big(X_{1} - \E\big[X_1|W_2,Y_1\big]\Big)^2\Big], \label{eq:MMSEweak}
\end{align}
where 
\begin{align}
\Q_1(P) &= \bigg\{  \QQ_{U_1W_2|X_0} \;\text{ s.t. }\; P = E_{\QQ}\big[U_{1}^2\big],\nonumber \\
& I(U_1;Y_1|W_2,X_0) - I(W_2;X_0)\geq0,  \nonumber \\
& (X_0,U_1,W_2,X_1,Y_1, U_2)\;  \text{ are continuous }\bigg\},
\end{align}
The random variables $(X_0,W_2,U_1)$ drawn according to $\QQ'_{U_1W_2|X_0}$ which is optimal for \eqref{eq:MMSEweak}, have covariance matrix
\begin{align}
K = 
\begin{pmatrix}
Q & \rho_1\sqrt{QV} & \rho_2\sqrt{QP} \\
\rho_1\sqrt{QV} &V & \rho_3\sqrt{VP} \\
 \rho_2\sqrt{QP} &  \rho_3\sqrt{VP} & P\\
\end{pmatrix},\label{eq:CovarianceMatrix}
\end{align}
where the correlation coefficients $(\rho_1,\rho_2,\rho_3)\in[-1,1]^3$ are such that $Q V  P \cdot \big(1- \rho_1^2- \rho_2^2- \rho_3^2 + 2 \rho_1 \rho_2\rho_3\big)\geq0$, i.e. $K$ is semi-definite positive.

We denote by $\QQ_{U_1W_2|X_0}$ the Gaussian conditional distribution such that $(X_0,W_2,U_1)\sim  \mc{N}(0,K)$. According to \cite[Maximum Differential Entropy Lemma, pp. 21]{EK11nit}, we have
\begin{align}
&\E_{\QQ'}\Big[\Big(X_{1} - \E\big[X_1|Y_1,W_2\big]\Big)^2\Big]
\geq  \frac{1}{2\pi e} \cdot 2^{2 h(X_1|Y_1,W_2)}\label{eq:InequalityMMSE1}\\
 &=\E_{\QQ}\Big[\Big(X_{1} - \E\big[X_1|Y_1,W_2\big]\Big)^2\Big].\label{eq:InequalityMMSE2}
\end{align}
Moreover, both $\PP_{X_0}\QQ'_{U_1W_2|X_0}$ and $\PP_{X_0}\QQ_{U_1W_2|X_0}$ satisfy \begin{align}
0\leq& I_{\QQ'}(U_1;Y_1|W_2,X_0) - I_{\QQ'}(W_2;X_0) \\
= &h_{\QQ'}(Y_1|W_2,X_0)- h_{\QQ}(Y_1|W_2,X_0,U_1) \nonumber\\&- h_{\QQ}(X_0) + h_{\QQ'}(X_0|W_2) \label{eq:MajorationGaussian1}\\
 \leq &h_{\QQ}(Y_1|W_2,X_0)- h_{\QQ}(Y_1|W_2,X_0,U_1) \nonumber\\&- h_{\QQ}(X_0) + h_{\QQ}(X_0|W_2) \label{eq:MajorationGaussian2}\\
= &I_{\QQ}(U_1;Y_1|W_2,X_0) - I_{\QQ}(W_2;X_0),
\end{align}
where \eqref{eq:MajorationGaussian1} comes from $h_{\QQ'}(X_0)=h_{\QQ}(X_0)$ and $h_{\QQ'}(Y_1|W_2,X_0,U_1)= h_{\QQ}(Z)=h_{\QQ}(Y_1|W_2,X_0,U_1)$, and  \eqref{eq:MajorationGaussian2} comes from \cite[(8.92), pp. 256]{CT06eoit}.

\begin{lemma}\label{lemma:EntropyComputationRho}
Assume that $(X_0,W_2,U_1)\sim  \mc{N}(0,K)$, then
\begin{align}
&I(U_1;Y|X_0,W_2)  - I(X_0;W_2) \nonumber\\
&=  \frac{1}{2 }\log_2 \bigg( \frac{ P}{ N}  \cdot (1- \rho_1^2 - \rho_2^2 - \rho_3^2 + 2 \rho_1\rho_2\rho_3) + (1- \rho_1^2)\bigg),\label{eq:IC}\\
&\E_{\QQ}\Big[\Big(X_{1} - \E\big[X_1|Y_1,W_2\big]\Big)^2\Big] \nonumber\\
&=  \frac{N\cdot \Big(Q \cdot (1-\rho_1^2 )+  P \cdot (1-\rho_3^2)+ 2\sqrt{QP} \cdot (\rho_2 -  \rho_1 \rho_3)\Big)}{N + \Big( Q\cdot (1-\rho_1^2) +  P\cdot (1-\rho_3^2) +  2\sqrt{QP} \cdot (\rho_2 -  \rho_1 \rho_3) \Big)}.\label{eq:MMSE}
\end{align}
\end{lemma}
The proof of Lemma \ref{lemma:EntropyComputationRho} is stated in Sec. \ref{sec:ProofLemmaEntropyComputation}.
Note that the equations \eqref{eq:IC} and \eqref{eq:MMSE} do not depend on the variance parameter $V$ of the auxiliary random variable $W_2$. Also, when \eqref{eq:IC} is positive the matrix $K$ is semi-definite positive.

By using Lemma \ref{lemma:EntropyComputationRho}, we reformulate \eqref{eq:MMSEweak} and since the function $x\to \frac{N\cdot x}{N+ x}$ is strictly increasing for all $x\geq0$, the optimal parameters $(\rho_1^{\star}, \rho_2^{\star}, \rho_3^{\star})\in[-1,1]^3$ minimize
\begin{align}
Q \cdot (1-\rho_1^2 )+  P \cdot (1-\rho_3^2)+ 2\sqrt{QP} \cdot (\rho_2 -  \rho_1 \rho_3) ,\label{eq:ConverseMinimizationCriteriaPb22}
\end{align}
under the constraint
\begin{align}
&\frac{ P}{ N}  \cdot (1- \rho_1^2 - \rho_2^2 - \rho_3^2 + 2 \rho_1\rho_2\rho_3) - \rho_1^2  \geq 0\label{eq:ConverseMinimizationPb2}\\
\Longleftrightarrow & (1- \rho_1^2)\cdot (1  - \rho_3^2) -  \frac{N}{P}  \cdot \rho_1^2 \geq   (\rho_2 -  \rho_1\rho_3)^2   ,
\end{align}
which yields
\begin{align}
\rho_2^{\star}  =&   \rho_1\rho_3  - \sqrt{(1- \rho_1^2)\cdot (1  - \rho_3^2) -  \frac{N}{P}  \cdot \rho_1^2}.\label{eq:Rho2}
\end{align}

\begin{lemma}\label{lemma:OptimalRho1Rho3}
If $Q> 4N$ and $P\in[P_1,P_2]$, then 
\begin{align}
{\rho_1^{\star}}^2= \frac{P\cdot Q - (P+N)^2}{(P+N)\cdot Q},\qquad
{\rho_3^{\star}}^2=0.
\end{align}
If $Q\leq4N$ \emph{or} if $Q> 4N$ and $P\in[0,P_1]\cup[P_2,Q]$, then 
\begin{align}
{\rho_1^{\star}}^2=0,\qquad
{\rho_3^{\star}}^2=0.
\end{align}
If $P> Q$, then 
\begin{align}
{\rho_1^{\star}}^2=0,\qquad
{\rho_3^{\star}}^2=\frac{P-Q}{P}.
\end{align}
\end{lemma}
The proof of Lemma \ref{lemma:OptimalRho1Rho3} is stated in Sec. \ref{sec:ProofLemmaOptimalRho}.
We obtain the lower bound by replacing the optimal parameters $(\rho_1^{\star}, \rho_2^{\star}, \rho_3^{\star})\in[-1,1]^3$ in \eqref{eq:MMSE}, that is if $P\in[0,Q]$
\begin{equation}
S_{\textsf{c}}(P) \geq
\begin{cases}
\frac{N\cdot(Q-N-P)}{Q} & \text{if } Q> 4N \text{ and } P\in[P_1,P_2],\\
\frac{ \big(\sqrt{Q}-  \sqrt{P} \big)^2 \cdot N}{ \big(\sqrt{Q}-  \sqrt{P} \big)^2 + N}& \text{otherwise. }
\end{cases}\label{eq:SolutionContinuousConverse}
\end{equation}


\subsection{Upper bound for the Theorem \ref{theo:CharacterizationContinuousRV}}\label{sec:AchievabilityProof}

\subsubsection{Linear Scheme}
By considering the best linear scheme 
\begin{align}
U_1 =& 
\begin{cases}
- \sqrt{\frac{P}{Q}} \cdot X_0 ,\quad &\text{ if } P<Q,\\
 - X_0 +\sqrt{P-Q},\quad &\text{ if } P\geq Q,
 \end{cases}
\end{align}
we have that $S_{\textsf{c}}(P) \leq S_{\ell}(P) $, for all $P\geq0$.

\subsubsection{Case where $Q> 4N$ and $P\in[P_1,P_2]$}

The upper bound of Theorem \ref{theo:CharacterizationContinuousRV} can be obtained by using a time sharing strategy between the two linear schemes with parameters $P_1$ and $P_2$. We obtain the same result by assuming that the random variables $(U_1,W_1,W_2)$ are drawn according to 
\begin{align}
&W_2 = \sqrt{\frac{P+N}{PQ - (P+N)^2}}  \cdot \Big( X_0 - Z_0 \Big)\sim \mathcal{N}\Big(0,1\Big),\nonumber\\
&\text{ with } Z_0 \sim \mathcal{N}\bigg(0, \frac{QN + (P+N)^2}{P+N}\bigg) \text{ and } Z_0 \perp W_2,\label{eq:AchievabilityW2}\\\
&W_1= \frac{PQ -(P+N)^2}{Q(P+N)} \cdot  X_0+U_0, \text{ with }U_0\perp (X_0,W_2)
\nonumber\\
&\text{ and } U_0 \sim   \mathcal{N}\bigg(0, \frac{N\cdot\big(PQ -(P+N)^2\big)}{QN + (P+N)^2}\bigg) ,\label{eq:AchievabilityW1}\\
&U_1 = - \frac{(P+N)^2}{QN + (P+N)^2} \cdot X_0  + U_0
\nonumber\\
&+  \sqrt{\frac{PQ - (P+N)^2}{P+N}}\cdot  \frac{(P+N)^2}{QN + (P+N)^2} \cdot W_2.
\end{align}
Then we have
\begin{align}
& I(W_1;Y_1,W_2) - I(W_1;X_0,W_2)  =  I(U_1;Y_1|X_0,W_2)  \\
 &= I(X_0;W_2)=  \frac{1}{2 }\log_2\bigg(1 +   \frac{ PQ -(P+N)^2}{QN + (P+N)^2}\bigg),
 \end{align}
and 
 \begin{align}
S_{\textsf{c}}(P) \leq& \frac{N\cdot (Q - P - N)}{Q}.\label{eq:coordinationMMSEachie}
\end{align}

\subsection{Proof of Lemma \ref{lemma:EntropyComputationRho}}\label{sec:ProofLemmaEntropyComputation}

We consider  $(X_0,W_2,U_1)\sim \mathcal{N}(0,K)$ with $K$ defined in \eqref{eq:CovarianceMatrix}, which together with \eqref{eq:Gaussian3}, induces the  Gaussian random variables $(X_0,W_2,Y_1)$ whose entropy is
\begin{align}
&h(X_0,W_2,Y) =\frac{1}{2 }\log_2 \bigg((2\pi e)^3\cdot Q V \\
&\times  \Big( P \cdot (1- \rho_1^2 - \rho_2^2 - \rho_3^2 + 2 \rho_1\rho_2\rho_3) + N \cdot (1- \rho_1^2)  \Big) \bigg).
\end{align}
Therefore we have
\begin{align}
&I(U_1;Y|X_0,W_2)  - I(X_0;W_2)\nonumber\\
=& h(X_0,W_2,Y)  - h(Y|U_1,X_0,W_2) - h(X_0) - h(W_2)\\
=&  \frac{1}{2 }\log_2 \bigg( \frac{ P}{ N}  \cdot (1- \rho_1^2 - \rho_2^2 - \rho_3^2 + 2 \rho_1\rho_2\rho_3) + (1- \rho_1^2)\bigg). 
\end{align}
According to \eqref{eq:Gaussian1} and \eqref{eq:Gaussian3} the entropy of $(X_1,W_2,Y_1)$ writes
\begin{align}
&h(X_1,W_2,Y) = \frac{1}{2 }\log_2 \bigg((2\pi e)^3\cdot   V \cdot N\\
&\times \Big(Q \cdot (1-\rho_1^2 )+  P \cdot (1-\rho_3^2)+ 2\sqrt{QP} \cdot (\rho_2 -  \rho_1 \rho_3)\Big)\bigg),
\end{align}
and hence 
\begin{align}
&\E\Big[\big(X_1 - \E(X_1|Y,W_2)\big)^2\Big] \nonumber\\
=&  \frac{N\cdot \Big(Q \cdot (1-\rho_1^2 )+  P \cdot (1-\rho_3^2)+ 2\sqrt{QP} \cdot (\rho_2 -  \rho_1 \rho_3)\Big)}{N + \Big( Q\cdot (1-\rho_1^2) +  P\cdot (1-\rho_3^2) +  2\sqrt{QP} \cdot (\rho_2 -  \rho_1 \rho_3) \Big)}.
\end{align}

\subsection{Proof of Lemma \ref{lemma:OptimalRho1Rho3}}\label{sec:ProofLemmaOptimalRho}

We replace $\rho_2^{\star}$ in \eqref{eq:ConverseMinimizationCriteriaPb22} and we define \begin{align}
&f(\rho_1^2,\rho_3^2) = Q \cdot (1-\rho_1^2 )+  P \cdot (1-\rho_3^2)\nonumber\\
&- 2\sqrt{QP} \cdot  \sqrt{(1- \rho_1^2)\cdot (1  - \rho_3^2) -  \frac{N}{P}  \cdot \rho_1^2}.
\end{align}
Note that $f$ is well defined if $\rho_1^2\leq \frac{P}{P+N}$ and $\rho_3^2 \leq 1 -   \frac{N}{P}  \cdot \frac{\rho_1^2}{1- \rho_1^2}$. 
\begin{align}
&\frac{\partial f(\rho_1^2,\rho_3^2)}{\partial \rho_3^2} = \sqrt{PQ}\cdot \frac{1-\rho_1^2}{\sqrt{(1- \rho_1^2)\cdot (1  - \rho_3^2) -  \frac{N}{P}  \cdot \rho_1^2}} - P,
\end{align}
then for all $\rho_1^2\leq \frac{P}{P+N}$, the optimal ${\rho_3^2}^{\star}(\rho_1^2)$ is
\begin{align}
{\rho_3^2}^{\star}(\rho_1^2) = \max\Bigg(1  -  \bigg( \frac{Q}{P}\cdot \Big(1-\rho_1^2\Big)+ \frac{N}{P}  \cdot \frac{\rho_1^2}{1- \rho_1^2}\bigg),0\Bigg).\label{eq:OptimalRho33}
\end{align}
We introduce the parameters
\begin{align}
\rho_a =&  \frac{2Q -(P+N) - \sqrt{(P+N)^2 - 4QN} }{2Q}, \label{eq:Condition2LemmaRho3}\\
\rho_b =&  \frac{2Q -(P+N) +\sqrt{(P+N)^2 - 4QN} }{2Q},\label{eq:Condition3LemmaRho3}
\end{align}
and we define the function
\begin{align}
&F(\rho_1^2)=f\Big(\rho_1^2,{\rho_3^2}^{\star}(\rho_1^2) \Big)\nonumber\\
=&\begin{cases}
Q \cdot (1-\rho_1^2 )+  P  \\
\qquad- 2\sqrt{QP} \cdot  \sqrt{1- \rho_1^2\cdot \frac{P+N}{P}}  &\text{ if }  0  \leq \rho_1^2\leq  \rho_a ,\\
N  \cdot \frac{\rho_1^2}{1- \rho_1^2}  & \text{ if }   \rho_a  \leq  \rho_1^2\leq\rho_b,\\
Q \cdot (1-\rho_1^2 )+  P  \\
\qquad- 2\sqrt{QP} \cdot  \sqrt{1- \rho_1^2\cdot \frac{P+N}{P}} &\text{ if }  \rho_b \leq \rho_1^2\leq \frac{P}{P+N}.
\end{cases}
\end{align}
The function $F(\rho_1^2)$ is continuous in  $ \rho_a$ and $\rho_b$. We define 
\begin{align}
\rho^{\star} =&  \frac{P\cdot Q - (P+N)^2}{(P+N)\cdot Q}.
\end{align}
$\bullet$ If $Q> 4N$ and $P\in[P_1,P_2]$, then the function $F(\rho_1^2)$ is decreasing over the interval $\rho_1^2\in[0,\rho^{\star}]$ and increasing over the interval $\rho_1^2\in[\rho^{\star},\frac{P}{P+N}]$, then the optimal parameters are
\begin{align}
\rho_1^2= \rho^{\star},\qquad \rho_3^2= 0.
\end{align}
$\bullet$ If $Q\leq4N$ or if $Q> 4N$ and $P\in[0,P_1]\cup[P_2,Q]$, then the optimal parameters are $\rho_1^2=\rho_3^2=0$.\\
$\bullet$ If $P>Q$, then 
\begin{align}
\rho_1^2= 0,\qquad \rho_3^2=\frac{P-Q}{P} .
\end{align}



\begin{thebibliography}{1}

\bibitem{W68acis}
H.~Witsenhausen, ``A counterexample in stochastic optimum control,'' \emph{SIAM Journal on Control}, vol.~6, no.~1, pp. 131--147, 1968.

\bibitem{YB13sncs}
S.~Y\"uksel and T.~Basar, \emph{{Stochastic Networked Control Systems: Stabilization and Optimization under Information Constraints}}, ser. Systems
  \& Control Foundations \& Applications.\hskip 1em plus 0.5em minus
  0.4em\relax New York, NY: Springer, 2013.

\bibitem{TT17alsa}
S.-H.~Tseng, A.~Tang. ``A Local Search Algorithm for the Witsenhausen’s Counterexample.,'' in \emph{Proc. IEEE CDC}, Dec. 2017.

\bibitem{KGOS11iscc}
J.~Karlsson, A.~Gattami, T.~J.~Oechtering, and M.~Skoglund, ``Iterative source-channel coding approach to Witsenhausen’s
counterexample,'' in \emph{Proc. IEEE ACC,} 2011.

\bibitem{WV11wcav} 
Y.~Wu and S.~Verdu, ``Witsenhausen’s counterexample: A view from optimal transport theory,'' in \emph{Proc. IEEE CDC,} 2011.

\bibitem{MH15ofam}
W.~M.~McEneaney and S.~H.~Han, ``Optimization formulation and monotonic solution method for the Witsenhausen problem,'' \emph{Automatica,} 2015.

\bibitem{GSV05miam}
D.~Guo, S.~Shamai and S.~Verdu, ``Mutual information and minimum mean-square error in Gaussian channels,'' in \emph{IEEE Trans.  Inform. Theory,} April 2005.

\bibitem{GS10wcaa}
P.~Grover  and  A.~Sahai, ``Witsenhausen's  counterexample  as  assisted interference  suppression,'' in \emph{Int.  J.  Syst.,  Control  Commun.},  
2010.

\bibitem{CM12owct}
C.~Choudhuri and U.~Mitra, ``On Witsenhausen's counterexample: the asymptotic vector case,'' \emph{2012 IEEE ITW}, pp. 162--166, July 2012.

\bibitem{MO18ocdv}
M.~Le~Treust and T.~J.~Oechtering, ``{Optimal Control Designs for Vector-valued Witsenhausen Counterexample Setups},''
  \emph{IEEE 56th Annual Allerton Conf. on Commun., Control, and Comp.}, Sept. 2018.

\bibitem{MT17jeco}
M.~Le.~Treust, ``{Joint Empirical Coordination of Source and Channel},''
  \emph{IEEE Trans. Inf. Theory}, vol.~63, no.~8, pp. 5087--5114, Aug. 2017.

 \bibitem{OL19cccd}
T.~J.~Oechtering and M.~Le~Treust, ``{Coordination Coding with Causal Decoder for Vector-valued Witsenhausen Counterexample Setups},''
  \emph{IEEE Information Theory Workshop (ITW)}, Aug. 2019.

\bibitem{VOS18hiwp}
M.~T. Vu, T.~J. Oechtering, and M.~Skoglund, ``{Hierarchical Identification With Pre-Processing},'' 
  \emph{IEEE Trans. Inf. Theory}, Jan. 2020.


\bibitem{CT06eoit}
T.~M. Cover and J.~A. Thomas, \emph{Elements of Information Theory},
  2nd~ed.\hskip 1em plus 0.5em minus 0.4em\relax Wiley \& Sons, 2006.


\bibitem{EK11nit}
A.~El~Gamal and Y.-H. Kim, {\em Network Information Theory}.
\newblock Cambridge University Press, Dec. 2011.












\end{thebibliography}
\end{document}